\documentclass{superfri}

\usepackage[utf8]{inputenc}

\usepackage{xspace}
\def\CO2 {CO$_2$\ }
\newcommand\dslash{\ensuremath{{\not}D}}

\newcommand{\clqcd}{CL\kern-.25em\textsuperscript{2}QCD}

\def\LOEWECSC{\mbox{LOEWE-CSC}}
\def\SANAM{\mbox{Sanam}}
\newcommand{\LCSC}{\mbox{L-CSC}}

\def\splitfirstchar#1#2\sentinel{\textbf{#1}#2}

\newcommand*{\eg}{e.\,g.\@\xspace}
\newcommand*{\ie}{i.\,e.\@\xspace}

\begin{document}

\author{D.~Rohr\footnote{\label{fias}Frankfurt Institute for Advanced Studies, Goethe University Frankfurt, Department for High Performance Computing, Ruth-Moufang-Str. 1, 60438 Frankfurt, Germany, rohr@compeng.uni-frankfurt.de},
G.~Nešković\footnoteref{fias},
V.~Lindenstruth\footnoteref{fias}${}^,$\footnote{GSI Helmholtz Center for Heavy Ion Research, Planckstraße 1, 64291 Darmstadt, Germany}}

\title{The L-CSC cluster: Optimizing power efficiency to become the greenest supercomputer in the world in the Green500 list of November 2014}

\maketitle{}

\begin{abstract}
The L-CSC (Lattice Computer for Scientific Computing) is a general purpose compute cluster built with commodity hardware installed at GSI.
Its main operational purpose is Lattice QCD (LQCD) calculations for physics simulations.
Quantum Chromo Dynamics (QCD) is the physical theory describing the strong force, one of the four known fundamental interactions in the universe.
L-CSC leverages a multi-GPU design accommodating the huge demand of LQCD for memory bandwidth.
In recent years, heterogeneous clusters with accelerators such as GPUs have become more and more powerful while supercomputers in general have shown enormous increases in power consumption making electricity costs and cooling a significant factor in the total cost of ownership.
Using mainly GPUs for processing, L-CSC is very power-efficient, and its architecture was optimized to provide the greatest possible power efficiency.
This paper presents the cluster design as well as optimizations to improve the power efficiency.
It examines the power measurements performed for the Green500 list of the most power-efficient supercomputers in the world which led to the number~1 position as the greenest supercomputer in November 2014.

\keywords{L-CSC, HPL, Linpack, Green500, GPU, Energy Efficiency, HPC, LQCD}
\end{abstract}

\section*{Introduction}

Quantum Chromo Dynamics (QCD) is the physical theory of the strong force, which describes the interaction between quarks and gluons, the fundamental constituents of hadronic matter in the universe.
It is a highly nonlinear theory where perturbative methods are only applicable in a small regime.
Lattice QCD (LQCD) uses a discretization in a space time grid, and it is the only general \textit{a priory} approach to QCD computations.
LQCD requires the inversion of the Dirac operator, which is usually performed by a conjugate gradient algorithm, which involves a sparse matrix-vector-multiplication called \dslash{}.
This \dslash{} operator is the computational hotspot of LQCD applications and therefore is responsible for a majority of the runtime of the program.
The bottleneck in \dslash{} is usually not the compute performance but the memory bandwidth, because sparse matrix-vector-multiplications require many memory loads per compute operation compared to other matrix operations with dense matrices like DGEMM.
Hence, for a compute cluster with LQCD as primary focus, a large memory bandwidth is paramount.

Supercomputers are inevitable in today's research.
Scientific challenges demand the fastest possible supercomputers, but it is prohibitively expensive to acquire more and more compute power through the use of more and more electricity.
In order to use the available resources to the maximum, computers have to become more power-efficient.
During the last several years, heterogeneous HPC clusters combining traditional processors with special accelerators such as GPUs or the Xeon Phi have been proven to deliver both superior compute performance and energy efficiency.
In an effort to raise awareness for power efficiency, the Green500 list~\cite{Sharma2006} provides a list of supercomputer power efficiencies and presents the ``greenest'' supercomputers in the world.

This paper presents L-CSC (Lattice Computer for Scientific Computing), which is built with commodity hardware and features four high-performance GPUs per compute node.
It is organized as follows:
Section~\ref{sec:cluster} describes the hardware of the cluster and why it is suited for LQCD.
The section outlines the design decisions for good power efficiency.
Section~\ref{sec:linpack} illustrates some optimizations we applied to achieve the best efficiency in the Linpack benchmark.
Finally, section~\ref{sec:green500} describes the efforts required to obtain an accurate and reasonable power measurement for the Green500 list and presents the results.

\section{The L-CSC cluster}
\label{sec:cluster}

In order to access a broad variety of hardware and to reduce acquisition costs, L-CSC is based on off-the-shelf components.
Its design follows the LOEWE-CSC and Sanam~\cite{bib:sanampaper} clusters, which have proven the validity of the commodity hardware approach for GPU accelerated HPC clusters.
L-CSC is a general purpose cluster that can run any kind of software, although its main focus is LQCD.

\label{sec:related}

\LCSC{} continues a trend of increasing performance and memory density of compute nodes as set by its predecessors, \LOEWECSC{} and \SANAM{}.
Table~\ref{node_trend} illustrates this trend.
Increased memory size enables larger HPC tasks to be executed on a single node and increased processing power shortens the wall time.
Consequently, this reduces the number of nodes and the size of the network in the cluster, which reflects positively on power efficiency and acquisition cost.

\tab{node_trend}{Comparison of \LOEWECSC{}, \SANAM{}, and \LCSC{} nodes (all numbers are aggregate values per compute node)}{
  \begin{tabular}{lccc}
    \hline
\textbf{Component}~~~    & \textbf{\LOEWECSC{}} & \textbf{\SANAM{}} & \textbf{~~~\LCSC{}~~~} \\\hline
CPU cores & 24 & 32 & 40 \\
GPUs & 1 & 4 \small{(2x dual-GPU)} & 4\\
System memory & 64~GB & 128~GB & 256~GB\\
GPU stream processors & 1600 & 7168 & 11264\\
GPU memory & 1~GB & 12~GB & 64~GB\\
GPU peak memory bandwidth & 153.6~GB/s & 960~GB/s & 1280~GB/s\\
Peak Performance [fp64 GFLOPS] & 745.6 & 3661 & 10618\\\hline
\end{tabular}
}

The most important criteria for a supercomputer with LQCD-focus are memory bandwidth and memory capacity.
Memory bandwidth defines the compute performance and memory capacity defines the maximum lattice size.
The performance of \dslash{} depends more or less linearly on the memory bandwidth and it is possible to use a large fraction of the theoretically available bandwidth in the application (Bach et al.~\cite{pos} show more than 100~GFLOPS which translates to about 80\% of the peak memory bandwidth with the OpenCL application employed on L-CSC).
The demands with respect to memory capacity are a bit more complex.
It is mandatory that the lattice fits in GPU memory.
If it does fit, no additional memory can be used at all.
Hence, memory should not be chosen too large in the first place.
For L-QCD calculations, the extent of the time dimension of the lattice is anti-proportional to the temperature.
Thermal lattices ($T > 0$) need much less memory than lattices with~$T \approx 0$.
As a different aspect, the distance of the lattice points can be decreased for better accuracy requiring more memory, but this also slows down the program.
Hence, the answer to the question of how much memory is needed depends on the actual problem.
A memory of~$3$ GB is already enough for most thermal lattice sizes~($T > 0$)~\cite{bib:sanampaper}, but has some limitations.
By and large, we consider~$16$~GB of L-CSC's S9150 cards sufficient for almost all lattices.

To make things even more complex, one can distribute the lattice over multiple GPUs or even over different compute nodes.
Tests on the Sanam cluster have shown a performance decrease on the order of~20\%, when more than one GPU is used.
The paradigm for L-CSC is to run most lattices on a single-GPU only, while there is still the possibility of using multiple GPUs for very large ones.
Still, multiple GPUs inside a compute node can be fully used in parallel to compute independent lattices.
Since LQCD needs a lot of statistic, involving a great deal of lattices, this approach is very efficient.

Overall, the design goal was four GPU boards per node with maximum aggregate GPU memory bandwidth - under the constraint of sufficient memory per GPU.
Two GPU types have been chosen: The AMD FirePro S9150 GPU, featuring a capacity of~$16$~GB and a bandwidth of~$320$~GB/s.
And the AMD FirePro S10000 dual-GPU (i.e. eight GPU chips per node), with a capacity of~$2 \times 6$~GB ($6$~GB per GPU chip) and a bandwidth of~$2 \times 240$~GB/s, thus with a higher aggregate bandwidth than~S9150.
Besides the higher memory capacity, the S9150 has the additional advantage of being able to reduce the wall time for small jobs compared to the S10000 due to the higher per-GPU-chip bandwidth.
This is important for application development and testing, when a quick answer is needed.
L-CSC runs all larger lattices on the S9150, and the smaller latices on both S10000 and S9150.
Very large lattices can span multiple S9150 cards, having access to~$64$~GB of GPU memory per node.

L-CSC consists of 160 compute nodes with 48~S10000 GPUs and 592~S9150 GPUs.
Each compute node consists of an ASUS ESC4000 G2S/FDR server, two Intel \mbox{Ivy-Bridge-EP} ten-core CPUs, and 256~GB of DDR3-1600 memory.
In order to offer more flexibility for general purpose applications on the CPUs in parallel, two CPU models are used: 60 nodes have 3~GHz CPUs for applications with high CPU demands and 90 nodes have 2.2~GHz CPUs.
The interconnect is 56~GBit FDR InfiniBand with half bisectional bandwidth and fat-tree topology.
Our main OpenCL LQCD application is \clqcd{}.\footnote{https://github.com/CL2QCD/cl2qcd}
It achieves around~$135$~GFLOPS per S9150 GPU in \dslash{}, which is the core routing of LQCD, and the aggregate \dslash{} performance of the entire cluster is~$89.5$~TFLOPS~\cite{bib:lcsc}.
We had optimized it for the Sanam cluster and it performs very well on the new S9150 GPUs of L-CSC without additional modifications.
The theoretical peak performance of L-CSC of around~$1.7$ PFLOPS is in fact much higher than what we achieve in \clqcd{} because LQCD is memory bound~\cite{bib:lcsc}.

\section{Optimizing for best power efficiency in Linpack}
\label{sec:linpack}

The Linpack benchmark is the standard benchmark for measuring the performance of supercomputers.
The Green500 list presents the most power-efficient supercomputers in the world~\cite{Sharma2006}.
Its ranking is determined by the GFLOPS achieved in the Linpack normalized by the average electricity consumption during the Linpack run.

Even though L-CSC consists of commodity hardware, there are no unnecessary components that drain power.
The main contributors are the CPUs, GPUs, memory, chipset, network, and remote management.
Power consumption of the hard disk with scratch space in each node and of other components are comparatively small, given that each node features four GPUs with~$275$~W each.
Universal Serial Bus (USB) contributes significantly with up to~$20$~W.
L-CSC uses full USB suspend which amounts to the same savings as if USB were switched of completely, so USB does not play a role here.

Some additional optimizations boost L-CSC's power efficiency during the Linpack run for the Green500.
An InfiniBand-based network boot allows switching off hard disks, SATA controller, and all Ethernet LAN ports completely.
We have investigated the effects of hardware parameters such as fan speed as well as voltage and frequency of GPU and CPU on both power consumption and performance in detail.
Figure~\ref{power-perf-figure} shows some of our measurements.

\footnotetext{\label{fn:measurement2}
In the measurements for HPL performance, every measurement point corresponds to one compute node.
For each node, we have selected four GPUs of identical voltage ID and plugged these GPUs into the node, such that the GPU voltage on the~$x$-axis defines the voltage of each of the four GPUs in the node.
Consider that the~$x$-axis shows the voltage ID of the GPUs at~$900$~MHz.
Running at~$774$~MHz, the GPUs operate at a lower voltage.
For the~$x$-position of the~$774$~MHz measurements, we still use the voltage ID of the high frequency in order to identify the compute nodes.
($774$ and~$900$~MHz measurements of the same compute nodes are shown at the same~$x$-position.)
\stepcounter{footnote}
}

\footnotetext{\label{fn:measurement}
For these measurements, we always locked all settings to a fixed value (\eg deactivated power saving features and automatic fan speed adjustments) and used GPU clocks of~$774$~MHz to avoid throttling.
The workload is a continuous DGEMM loop.
For the power versus GPU fan speed measurement, we removed all GPUs from the servers to exclude GPU temperature effects and measure only the change in fan power.
The power v.s.~temperature curve is measured by letting the system heat up over a period of several minutes under load while the measurement is taken.
The GPU power consumption measurements for the right plot were performed on an ASUS ESC8000 server, the eight-GPU cousin of the L-CSC servers, to increase the fraction of the GPU power consumption in the total system power consumption.
\stepcounter{footnote}
}

\fig{width=\textwidth}{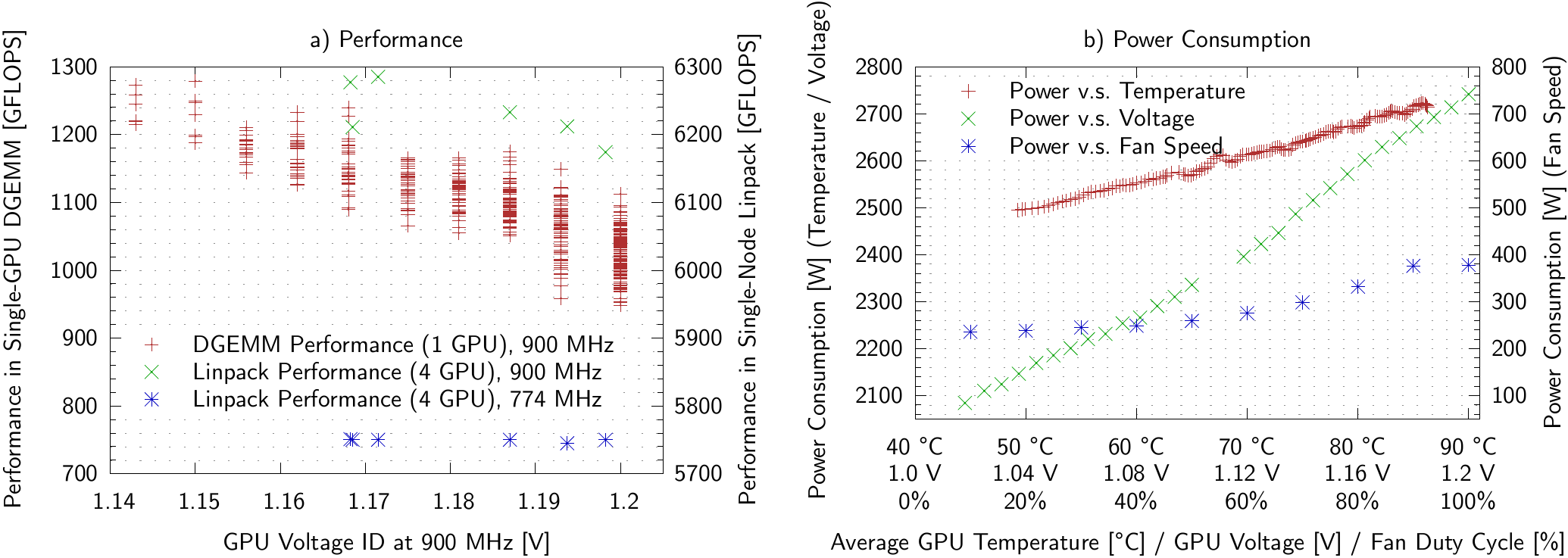}{{
Performance (a) and power (b) measurements of L-CSC nodes, S9150 GPUs, and system Fans.
Plot on the left shows the performance achieved in DGEMM (single-GPU) and HPL (single-node, \ie quad-GPU) at the stock clocks of~$900$~MHz and HPL performance at the most efficient clock rate of~$774$~MHz.
The $x$-axis represents the voltage of the employed GPUs at~$900$~MHz and it is obvious that the GPUs with higher voltage by trend throttle more and achieve less performance.\footnoteref{fn:measurement2}
Plot on the right shows how the power consumption of the full server varies with GPU temperature, GPU voltage, and FAN Speed.\footnoteref{fn:measurement}
See the footnotes for information on the measuring conditions.
}\label{power-perf-figure}}

\looseness=-1
Due to fluctuations in the manufacturing process every ASIC is a bit different and the vendors account for this by programming individual voltage IDs into their chips.
This means that every individual GPU runs only at the voltage its particular chip needs and especially different GPUs even of the same type operate at different voltages.
Today, CPUs and GPUs have a TDP limit and they will throttle their clock frequency under high load if their power consumption would exceed this limit otherwise.
Since GPUs operating at different voltages drain different power, the GPUs with lower voltage will hit this limit less frequently and hence operate on average at a higher frequency yielding better performance.
Figure~1a on the left visualizes this aspect in single-GPU DGEMM (matrix-matrix multiplication) and multi-GPU HPL benchmarks.
In the DGEMM case at 900~MHz, the GPUs that can operate at the lowest voltage of 1.1425V achieve an average DGEMM performance of~1250~GFLOPS compared to only~950 to~1100~GFLOPS for the slowest GPUs that operate at~1.2V.
HPL Performance at 900~MHz varies between~6175 and~6280~GFLOPS.
Because multi-node HPL distributes the workload evenly among all compute nodes, the slowest compute nodes dictate the performance.
In this case, it is very unfavorable if the compute nodes achieve different performances due to different GPU voltages.
Using a heuristic search in the parameter space of GPU voltage, GPU and CPU frequencies, fan speed settings, and settings for the HPL-GPU benchmark, we have identified the parameter set that we believe delivers the best power efficiency.
The optimal GPU frequency is~$774$~MHz.
Figure~1a (left) shows a completely flat performance profile for this energy-efficient configuration with~$774$~MHz, \ie no GPUs throttle and all nodes achieve the same performance.

An interesting observation in this context is that the highest clock rate of~$900$~MHz does not deliver the highest performance.
Due to the throttling, the GPU oscillates between the~$900$~MHz frequency and lower frequencies.
This is less efficient than constant operation at the highest possible frequency that does not throttle.
For instance, running with default GPU power management settings on L-CSC, we see higher constant DGEMM performance at a GPU frequency of~$820$~MHz than with~$900$~MHz.

The figure~1b (right) shows the dependency of the power consumption on fan speed, GPU voltage, and GPU temperature.
Obviously, the largest contribution by far comes from the GPU voltage.
For the final Linpack run, we have used the minimum voltage required for stable operation of all GPUs at the target frequency of~$774$~MHz.
Now, it is clear that one cannot operate at the lowest possible temperature and on low fan speeds at the same time because low fan speeds cause higher temperatures.
The power curve for different fan speeds shows a stronger slope for fan speeds above~$40\%$ and we have found~$40\%$ to be the optimum during the high-load phase of the Linpack benchmark.
Toward the end of a Linpack run, the load reduces significantly.
We account for this by employing a curve that defines different FAN duty cycles for different load levels / temperatures, which ensures that the FANs always run only at the minimum speed required.
This reduces further power consumption.

For running the Linpack benchmark we employed our HPL-GPU\footnote{https://github.com/davidrohr/hpl-gpu/wiki}~\cite{bib:isc} implementation of the benchmark, which we have developed and used for the LOEWE-CSC and Sanam clusters before.
It provides two operating modes: One optimized for maximum performance, and an alternative mode that sacrifices a small fraction of the performance to reduce the power consumption resulting in better net power efficiency.
This alternative efficiency-optimized mode was developed further for L-CSC and has been used for the Green500 result~\cite{bib:lcsc}.

\section{Measuring the power consumption for the Green500 list}
\label{sec:green500}

\looseness=-1
The Green500 ranking is determined by the quotient of the achieved performance in the Linpack benchmark divided by the average power consumption during Linpack execution.
Due to late installation of the system, only~56~nodes with S9150 GPUs were available for the Linpack benchmark in November 2014, which were connected by three InfiniBand switches in a ring-configuration.
We did not repeat the Linpack measurements on the full system which has gone in production operation meanwhile.
From scalability tests of HPL-GPU on the Sanam cluster and on subsets of L-CSC~\cite{bib:lcsc}, we assume the full system would achieve an almost identical power efficiency.
The current Green500 measurement methodology revision 1.2 is defined by~\cite{bib:eehpc}.
Tab.~\ref{levels} lists three measurement levels defined in this methodology document yielding different accuracies.

\tab{levels}{Measurement levels for Green500 with different accuracy}{
\begin{tabular}{llll}
\hline
Level & Components & Measured fraction of system & Duration \\
\hline
1 & Only compute nodes & At least~$\frac{1}{64}$ of the system & At least~20\% of the\\
& & & middle~80\% of the run \\
2 & Full cluster with network & At least~$\frac{1}{8}$ & Full runtime \\
& (network estimated) & & \\
3 & Full cluster with network & Full system & Full runtime \\
& (network measured) & & \\

\hline
\end{tabular}
}

Level~1 is provided for facilities without sufficient equipment for higher level measurements.
Unfortunately, the level~1 specifications are exploitable such that one can create measurements which show a higher power efficiency than actually achieved~\cite{bib:lcsc}.
Thus, higher levels are preferred.
The L-CSC installation had only one revenue grade power meter available (see \cite{bib:eehpc} for power meter requirements), and it was thus impossible to measure a larger fraction of the system at the accuracy required for level~2 or level~3.
Thus, only a level~1 measurement was feasible.
All measures were taken to make the result as accurate as possible.
Our measurement for L-CSC includes the entire Linpack run and we measured the entire cluster with the network.
Due to the lack of more revenue grade power measurement equipment, only two compute nodes could be measured.
However, in order to obtain the most accurate result, we have taken additional measures to mitigate the effect of measuring not the full system.
Power consumption variability of nodes can be estimated by measuring the efficiency of several individual nodes during single-node Linpack runs, which yielded the following values on seven randomly chosen nodes:

\vspace{3pt}
\textbf{5154.1, 5260.1, 5248.4, 5245.5, 5125.1, 5301.2, 5169.3~[MFLOPS/W].}
\vspace{3pt}

\noindent \looseness=-1
The results show a relatively small variation of only $\pm 1.2$\%.
In order to deliver the most accurate result, we used nodes with middle power consumption among the nodes we had measured individually before.
Hence, the difference to the full level three measurement is small.
The only aspect not fulfilling level~3 is the number of measured nodes.
With a deviation of less than~1.2\% between the nodes, and due to the fact that we have chosen average nodes for the final measument, we assume that our efficiency result is off by less than~$1\%$ from a full level three measurement.
(Surprisingly for us, the three switches only contribute with~$257$~W to the power consumption.)
In contrast, many other top ranked Green500 systems (\eg the ExaScaler systems and TSUBAME-KFC currently ranked on places 1, 2, 3, and 5 of the June 2015 list) do not take measures to ensure exact measurements but instead only measure a period with low power consumption according to level 1~\cite{bib:tsubame-kfc}.
We have shown that in case of L-CSC such a measurement overestimates the real efficiency by up to~$30$\%~\cite{bib:lcsc}, currently corresponding to the margin of the first over the fourth rank in the Green500 list.
This greatly deteriorates the comparability.
Accordingly, the newly released power measurement specification 2.0 RC 1~\cite{bib:eehpc2} for the Green500 list prohibit such course of action and will lead to better comparability of new measurements.

\section{Results and Conclusion}

The 56 nodes used for the measurement achieved a Linpack performance of~301.5~TFLOPS expending on average 57.2~kW and yielding an average efficiency of~5271.8~MFLOPS/W with a measurement error of less than~$1.2$\%.
With this result, L-CSC was awarded 1st place in the Green500 list of November 2014 as the most power-efficient supercomputer in the world.
We have selected lower clocks and voltages to achieve optimal performance.
The performance decrease is not very large in applications like Linpack that reach close to the peak performance because under such high load the GPUs cannot maintain the maximum clocks over a long time.
Essentially, when the GPU operates at its power limit, the achieved performance depends linearly on the power efficiency and we have seen that a slight decrease in clock speed can even lead to a better performance.
The energy efficiency improvements we observe are also applicable to our application.
Our LQCD \dslash{} kernel in particular is memory bound and little sensitive to frequency.
It suffers less than~$1.5$\% performance decrease with the efficiency-optimized settings.

\ack{We would like to thank Advanced Micro Devices, Inc. (AMD) and ASUSTeK Computer Inc. (Asus) for their support. A part of this work was funded by HIC for FAIR.}

\bibliography{citations}

\end{document}